\documentclass{article}

\usepackage{arxiv}

\usepackage[utf8]{inputenc} 
\usepackage[T1]{fontenc}    
\usepackage{hyperref}       
\usepackage{url}            
\usepackage{booktabs}       
\usepackage{amsfonts}       
\usepackage{nicefrac}       
\usepackage{microtype}      
\usepackage{lipsum}		
\usepackage{graphicx}
\usepackage[numbers]{natbib}
\usepackage{doi}
\usepackage{caption}
\usepackage{subcaption}
\usepackage{multirow}
\newcommand{\degree}{$^{\circ}$}

\title{Assessing the Privacy Risk of Cross-Platform Identity Linkage using Eye Movement Biometrics}

\author{ 
    \href{https://orcid.org/0000-0002-7656-2662}{\includegraphics[scale=0.06]{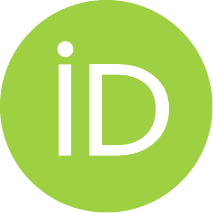}\hspace{1mm}Samantha Aziz} \\
	Department of Computer Science\\
	Texas State University\\
	San Marcos, TX 78666 \\
	\texttt{sda69@txstate.edu} \\
    \And	
    \href{https://orcid.org/0000-0001-7890-8842}{\includegraphics[scale=0.06]{orcid.pdf}
    \hspace{1mm}Oleg Komogortsev} \\
	Department of Computer Science\\
	Texas State University\\
	San Marcos, TX 78666 \\
	\texttt{ok@txstate.edu} \\
}

\date{}

\renewcommand{\headeright}{}
\renewcommand{\undertitle}{}

\hypersetup{
pdftitle={Assessing the Privacy Risk of Cross-Platform Identity Linkage using Eye Movement Biometrics},
pdfsubject={cs.HC},
pdfauthor={Samantha Aziz, Oleg Komogortsev},
pdfkeywords={eye tracking, privacy, reidentification, biometrics},
}

\begin{document}
\maketitle

\begin{abstract}
The recent emergence of ubiquitous, multi-platform eye tracking has raised user privacy concerns over re-identification across platforms, where a person is re-identified across multiple eye tracking-enabled platforms using personally identifying information that is implicitly expressed through their eye movement.
We present an empirical investigation quantifying a modern eye movement biometric model's ability to link subject identities across three different eye tracking devices using eye movement signals from each device.
We show that a state-of-the art eye movement biometrics model demonstrates above-chance levels of biometric performance (34.99\% equal error rate, 15\% rank-1 identification rate) when linking user identities across one pair of devices, but not for the other.
Considering these findings, we also discuss the impact that eye tracking signal quality has on the model's ability to meaningfully associate a subject's identity between two substantially different eye tracking devices.
Our investigation advances a fundamental understanding of the privacy risks for identity linkage across platforms by employing both quantitative and qualitative measures of biometric performance, including a visualization of the model's ability to distinguish genuine and imposter authentication attempts across platforms. 
\end{abstract}


\section{Introduction}
Eye tracking is becoming increasingly ubiquitous due to the recent large-scale integration of eye tracking functionality in mainstream consumer devices, including laptops and virtual reality headsets.
While eye tracking can enhance day-to-day activities by enabling functionalities such as foveated rendering~\cite{Patney2016} and intuitive gaze-based interaction~\cite{gaze-interaction}, it also raises concerns for user privacy.
Eye movements are produced by a complex interplay between the brain's physical structure and various neurological processes, and implicitly contain signatures of personal characteristics such as a user's identity, gender, and health status~\cite{Kroger2020}.
Because these traits are expressed unconsciously through eye movement, it is infeasible for individual users to voluntarily withhold this information from their eye movement signals.
This raises concerns for the potential of an adversary exploiting information encoded in eye movements to infer sensitive user characteristics.

Amid the growing discussions of privacy within the eye tracking research community~\cite{davidjohn2021privacy, Kroger2020, Li2021, Liebling2014,Liu2019, Steil2019a}, one potential new threat has emerged with the growing ubiquity of eye tracking devices: cross-platform identity linkage.
Because eye tracking is being employed in contexts involving laptops~\cite{Majaranta2014}, virtual/augmented reality~\cite{Patney2016}, and even vehicles~\cite{Singh}, a single person may utilize multiple eye tracking-enabled platforms throughout the course of their day.
In a setting where eye tracking is pervasive and ubiquitous, this person's identity may be linked across these platforms against their will using the personally identifying information encoded in their eye movement signals. 
This has both privacy implications for both individual users and security implications for systems that use eye movement biometrics as a means of access control.

Empirical investigations quantifying the risk of cross-platform identity linkage have been absent in literature because, until recently, the eye tracking research community lacked the data needed to support such an investigation. 
With the recent emergence of publicly available eye tracking data sets drawn from the same population of subjects~\cite{Aziz2022_hololens, Aziz2022, Lohr2023}, it is now possible to investigate the risk of cross-platform identity linkage using eye movement data.

This work explores the possibility that a user can be re-identified across eye tracking devices when using state-of-the-art eye movement biometrics (EMB) models. 
This study makes the following specific contributions:
\begin{enumerate}
    \item We present the first analysis of cross-platform identity linkage for EMB employing EyeKnowYouToo~\cite{Lohr2022_EKYT}, a state-of-the-art eye movement biometric authentication model, including a brief discussion of its privacy and security implications.
    \item We show that a modern EMB model can successfully link user identities across two of the three eye tracking platforms studied with above-chance accuracy during both biometric verification and identification.
    \item We use both quantitative and qualitative evaluations of biometric performance to discuss factors that affect an EMB model's ability to perform cross-platform re-identification, including the effect of eye tracking signal quality and the amount of data provided to the EMB model.
\end{enumerate}

\section{Background and Prior Work}

\subsection{Eye Movement Biometrics (EMB)}
First proposed as a biometric by Kasprowski and Ober~\cite{Kasprowski2004}, eye movement has emerged as an increasingly viable modality for biometric authentication. 
Because eye movements are produced by largely involuntary mechanisms of the oculomotor system, they are both person-specific and naturally resistant to spoofing.

Substantial research has been conducted exploring the viability of EMB to perform both verification~\cite{Holland2011_bio,Holland2013_bio,GEORGE2016_bio,Makowski2021, Lohr2022_EKYT, Lohr_eyeknowyou1} and identification~\cite{Bednarik2005_bio, Holland2013_bio, Lohr2022_EKYT, Jia2018, Makowski2021}.
Prior work in EMB distinguishes individuals by generating biometric templates based on features extracted from eye movement signals, such as average fixation duration, average saccade amplitude, and peak saccade velocity~\cite{Holland2011_bio,Holland2013_bio,GEORGE2016_bio} or on pupillary features such as pupil diameter~\cite{Bednarik2005_bio}. 
These EMB authentication models rely on careful extraction, fusion, and analysis of handcrafted features to reach verification error rates as low as 2.59\%~\cite{GEORGE2016_bio} and identification rates as high as 83.7\%~\cite{Holland2013_bio}.

State-of-the-art EMB models employ end-to-end deep learning~\cite{Makowski2021, Lohr2022_EKYT, Jia2018} to automatically extract features that meaningful for biometric authentication, rather than relying on static feature extraction and analysis.
These end-to-end approaches reliably achieve verification error rates below 4\%~\cite{Lohr2022_EKYT, Makowski2021} and identification rates exceeding 90\%~\cite{Lohr2022_EKYT} on eye movement data sets containing hundreds of unique subject identities.

Eye tracking signal quality metrics such as spatial accuracy, spatial precision, and sampling frequency affects eye movement biometric performance, as it is more difficult to extract meaningful biometric features from lower quality eye movement data.
While state-of-the-art approaches demonstrate some level of robustness to certain types of eye tracking signal quality degradation within a single device~\cite{PRASSE2020_bio, Lohr2022_EKYT}, it is unknown whether they are robust to differences in eye tracking signal quality when data from two different eye tracking devices is presented.
As a result, it is an open research question whether state-of-the-art EMB models can meaningfully re-identify the same users across multiple eye tracking platforms.

\subsection{Privacy Implications of Pervasive Eye Tracking}
Eye movements implicitly contain information about sensitive user attributes such as their identity, gender, age, and health status~\cite{Kroger2020}.
The primary privacy concern that we address in this work is whether a user's identity may be linked across these platforms using the eye movement data captured by their respective eye tracking devices.
The theoretical opportunity for an adversary to build such an association between devices is becoming increasingly likely due to the growing ubiquity of eye tracking functionality in mainstream computing devices.

This identity linkage can have implications for individual users.
For example, consider a user who uses both a laptop and a vehicle with eye tracking functionality.
If a remote adversary successfully links the user's identity between their laptop and their vehicle, they could exploit the car's GPS system to track their victim's movement.
Alternatively, it may be possible for an adversary to covertly capture a victim's eye movements in one device, then use it to bypass an EMB-based authentication mechanism in another system where the victim is enrolled.
This has larger security implications, as this authentication mechanism may be protecting sensitive information such as the victim's personal bank account or sensitive company information held in the victim's digital workplace platform.

Although the consequences of cross-platform identity linkage have been described both eye tracking literature~\cite{davidjohn2021privacy} and in general behavioral biometric research~\cite{Eberz2018}, the empirical risk of such violations occurring with eye tracking data specifically are underexplored.
Our work aims to contribute to a growing understanding of the privacy implications of pervasive eye tracking by presenting an exploration of cross-platform identity linkage using three eye tracking devices of varying signal quality.
In doing so, we aim to encourage the development of defense mechanisms that realistically address to user privacy concerns that emerge from ubiquitous, multi-platform eye tracking.

\section{Methodology}
\subsection{Threat Model}
\begin{figure}
    \centering
    \includegraphics[width=0.5\linewidth]{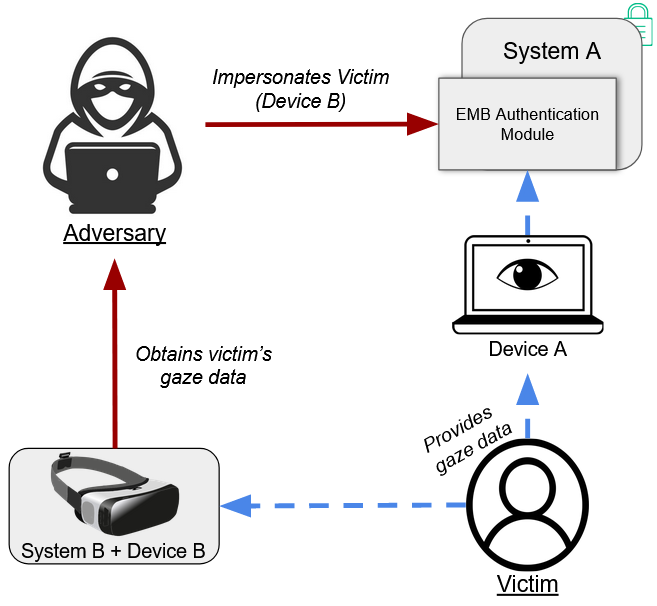}
    \caption{An example of an attack described by our threat model. Blue arrows indicate benign activity, and red arrows indicate malicious activity.}
    \label{fig:threat}
\end{figure}

This investigation concerns an attack scenario where an adversary attempts to bypass an EMB-based authentication system by presenting an enrolled user's eye movement data captured by a device other than the target system's eye tracking hardware.
Figure~\ref{fig:threat} shows an overview of this attack.

In this model, there exists a secure System A that employs EMB-based authentication as a form of access control, and cannot be accessed by any other means.
This authentication model accepts eye movement signals as input, which are represented as time-series data expressed as a sequence of $(x,y)$ coordinates respectively denoting the horizontal and vertical position of the user's gaze on the screen. The system does not employ auxiliary information such as images of the iris or periocular features for authentication.

When authenticating an enrolled user, the authentication module compares eye movement signals presented by the user to an enrollment template, and grants access if the two are sufficiently similar.
The authentication module operates as intended and cannot be modified or viewed by any means (e.g., the model cannot be re-trained or fine-tuned, its weights cannot be reverse-engineered, and enrollment templates cannot be viewed or altered).

We assume that a victim enrolled in System A are also enrolled in a second, less secure System B that incorporates eye tracking functionality with its own eye tracking device.
This system can either be an actual computing environment with eye tracking functionality whose insecurities can be exploited to obtain eye movement data (e.g., a victim's personal virtual reality headset) or an eye tracking data set used for research purposes that the adversary obtains from publicly available sources~\cite{davidjohn2023}.
In either case, an adversary obtains the victim's eye movement data captured by System B's eye tracking device.
The adversary then attempts to gain access to System A by presenting this stolen data to the System A's authentication module.
If the biometric features expressed in the victim's eye movement are sufficiently similar between each system's eye tracking device, the adversary is able use the victim's data obtained from System B to infiltrate System A under their identity.
Because the stolen eye movement data is both genuine and was never previously encountered by System A, mechanisms for detecting spoofing and replay attacks for eye movement are not sufficient to address this threat.

\subsection{Biometric Model}
We selected EyeKnowYouToo (EKYT), an end-to-end deep learning approach for EMB authentication, to serve as the secure biometric authentication module described in our threat model.
EKYT comprises an ensemble of four DenseNet-based convolutional neural networks trained to extract meaningful features for biometric authentication from a variety of eye movement tasks.
We chose this model because it is robust to template aging up to 37 months, which mitigates the potential impact of the large time interval exhibited by the data we use (described in Section~\ref{sec:data}) on the results of our current investigation.

For this investigation, we use pre-trained models from EKYT’s public release. 
These models were trained on eye movement data from an EyeLink 1000 eye tracking device that was temporally downsampled to 250~Hz.
Further details on EKYT's training methodology are described in~\cite{Lohr2022_EKYT}, and the pre-trained models employed in this study are publicly available at~\cite{Lohr2022_data}.

\label{sec:data}
\subsection{Data sets}
We employ eye movement data collected from 21 participants across three different eye tracking devices.
This data is taken from a subset of two publicly available eye tracking data sets: GazeBaseVR~\cite{Lohr2023} and SynchronEyes~\cite{Aziz2022}.
A summary of the eye tracking signal qualities exhibited by each data set is presented in Table~\ref{tab:data}.

GazeBaseVR contains eye movement data collected with a SensoMotoric Instruments (SMI) eye tracker embedded within a head-mounted HTC Vive virtual reality device at a 250 Hz sampling rate.
We employ eye movement data captured from the right eye during the TEX task collected during Round 1, Session 2, where participants completed a self-paced reading task.
The duration of this task varied between 27 and 66 seconds, with a mean reading time of 47 seconds.
The data that we use from GazeBaseVR will be referred to as data from the ``Vive'' device.

SynchronEyes contains eye movement data collected simultaneously in two eye trackers.
The first eye tracker is an EyeLink 1000 collecting monocular (left eye) data at a sampling rate of 1000 Hz, known hereon as the ``EyeLink'' data.
The second eye tracker is an AdHawk MindLink collecting monocular (right eye) data at a sampling rate of 500 Hz, known hereon as ``MindLink'' data. 
We employ reading data from the TEX task collected during Session 2, where eye movements were captured simultaneously in both eye tracking devices.
The duration of this reading task was 60 seconds for all participants.
For portions of this investigation, we also used eye movement data from the RAN task collected by the EyeLink during Session 2, where participants tracked a jumping dot.

GazeBaseVR and SynchronEyes were created approximately 23 months apart, which introduces potential effects of aging on biometric performance.
Aging effects on eye movements~\cite{spooner} can impact the temporal persistence of features that are informative for biometric authentication, which in turn affects the long-term reliability of a biometric system. 
EKYT's relative robustness against template aging effects makes it the most effective choice for mitigating the potential effects of aging that may be observed between GazeBaseVR recordings and SynchronEyes recordings, as the 23-month interval between the two data sets is well within the 37-month time frame explored in Lohr and Komogortsev's original work~\cite{Lohr2022_EKYT}.


\begin{table*}[]
    \centering
    \begin{tabular}{cccccl}
        \hline
        Name & Data set & Device & Spatial Accuracy & Spatial Precision & Task (Duration) \\
        \hline
        EyeLink & SynchronEyes & EyeLink 1000 & 0.63\degree & 0.09\degree & TEX, RAN (60 s) \\
        MindLink & SynchronEyes & AdHawk MindLink & 2.09\degree & 0.18\degree & TEX (60 s) \\
        Vive & GazeBaseVR & SMI Eye Tracker & 1.0\degree & 0.03\degree & TEX (27 - 67 s, mean 47 s) \\
        \hline
    \end{tabular}
    \caption{Summary of the eye movement data employed in this study. TEX refers to a reading task and RAN refers to a jumping dots task.}
    \label{tab:data}
\end{table*}
\label{sec:model}

\label{sec:preprocess}
\subsection{Data Preprocessing}
To enable use of the eye movement data we employ with EKYT, eye movement data were processed using the techniques described in Lohr and Komogortsev~\cite{Lohr2022_EKYT}.
Because we are employing a version of EKYT appropriate for 250~Hz sampling frequencies, data were first temporally downsampled to a sampling rate of 250 Hz.

We estimate velocity from the positional gaze data using a Savitzky-Golay differentiation filter~\cite{Savitzky1964} with order 2 and a window size of 7. 
This velocity data was then split into non-overlapping windows of 5 seconds. 
Within each window, we clamped velocities to physiologically feasible movement speeds of $\pm$1000\degree/sec to reduce the impact of noise, then z-scored velocity using the mean and standard deviation of eye movement velocities derived from EKYT's original training data and replaced NaN values with 0.
These processed velocity windows are then fed into the pre-trained EKYT model.
We note that, except for downsampling data to 250 Hz, all of this data processing would be completed by the biometric authentication module when a biometric sample is presented to the system for authentication; it would not be an adversary's responsibility to implement the data processing techniques described here.

\subsection{Evaluation Methodology}
EKYT transforms a sequence of eye movement windows into a 128-dimensional embedding vectors (known hereon as ``embeddings'') that represent subject identity.
We generate embeddings using 5 and 60 seconds' worth of data respectively for consistency with existing EMB literature~\cite{Makowski2021, Lohr2022_EKYT}.
We generate one embedding per recording in the data set, and associate it with both the device and the subject identity that produced the original eye movement sequence.

After generating these embeddings, we compute all pairwise similarity scores by calculating the cosine similarity between each embedding in the enrollment and the authentication sets.
These similarity scores are then used to determine biometric performance in both a verification and identification setting.

For biometric verification, the biometric system determines whether the embedding generated from the provided biometric sample is sufficiently similar to the enrollment template of the claimed identity.
Verification performance across the entire data set is evaluated using Equal Error Rate (EER), or the point at which the False Acceptance Rate and the False Rejection Rate are equal.
Because biometric verification only has two outcomes (i.e., the verification attempt is accepted or rejected), chance level performance is 50\% EER.

For biometric identification, the biometric system returns the identity of the subject whose embedding is most similar to the embedding of the biometric sample presented for authentication.
Identification performance is evaluated using rank-1 identification rate (IR), which measures how often the identities of the enrollment and authentication embeddings match.
For biometric identification, chance level performance is 1/N, where N is the number of enrolled users.

Because EKYT was trained on data from an EyeLink 1000 eye tracking device~\cite{Lohr2022_EKYT}, we use the embeddings created from the EyeLink portion of the data as enrollment templates.
Presenting data from the EyeLink for authentication simulates normal operation of the biometric system, and presenting data from the Vive or MindLink for authentication simulate an adversary's attempt to bypass the biometric system using the eye movement obtained from another eye tracking device. 
We assess biometric performance when presenting both ``legitimate'' EyeLink data and the ``stolen'' MindLink or Vive data for authentication.

\section{Results}
In each evaluation setting described herein, we use reading data from the EyeLink as the enrollment set.
We observe the biometric performance of the model when employing data from each device as the authentication set.
When using EyeLink data for both enrollment and authentication, we introduce additional data from SynchronEyes collected during a jumping dots task to serve as the authentication set. 
This evaluation serves as a baseline for comparison when authenticating with other devices.
Otherwise, all enrollment and authentication is done using a reading task taken from each device's respective data set. 

Verification and identification results are presented separately for 5-second and 60-second embeddings.
For this investigation, a chance-level verification rate is 50\% EER and a chance-level identification rate is $1/21 \approx 4.8\%$ IR.

\subsection{Cross-Platform Verification}
Table~\ref{tab:verification-results-5s} shows verification performance for each device when enrolling and authenticating on 5-second embeddings.
EKYT achieves a EER of 5.61\% when authenticating with EyeLink data, demonstrating that it can meaningfully distinguish between genuine and imposter identities when enrolling and authenticating on data from the same device.
In the context of our threat model, this means that the biometric authentication module achieves acceptably high biometric performance under normal, benign operating conditions.
EER deteriorates significantly when authenticating on data from non-EyeLink devices.
When using Vive data as the authentication set, EER degrades to 32.23\%, which is significantly worse but well above chance-level performance.
On the other hand, using MindLink data as the authentication set reduces biometric performance to chance-level performance of 49.13\%, indicating that MindLink embeddings are generally not meaningful for biometric verification in this setting.

\begin{figure}
    \centering
    \includegraphics[width=0.4\linewidth]{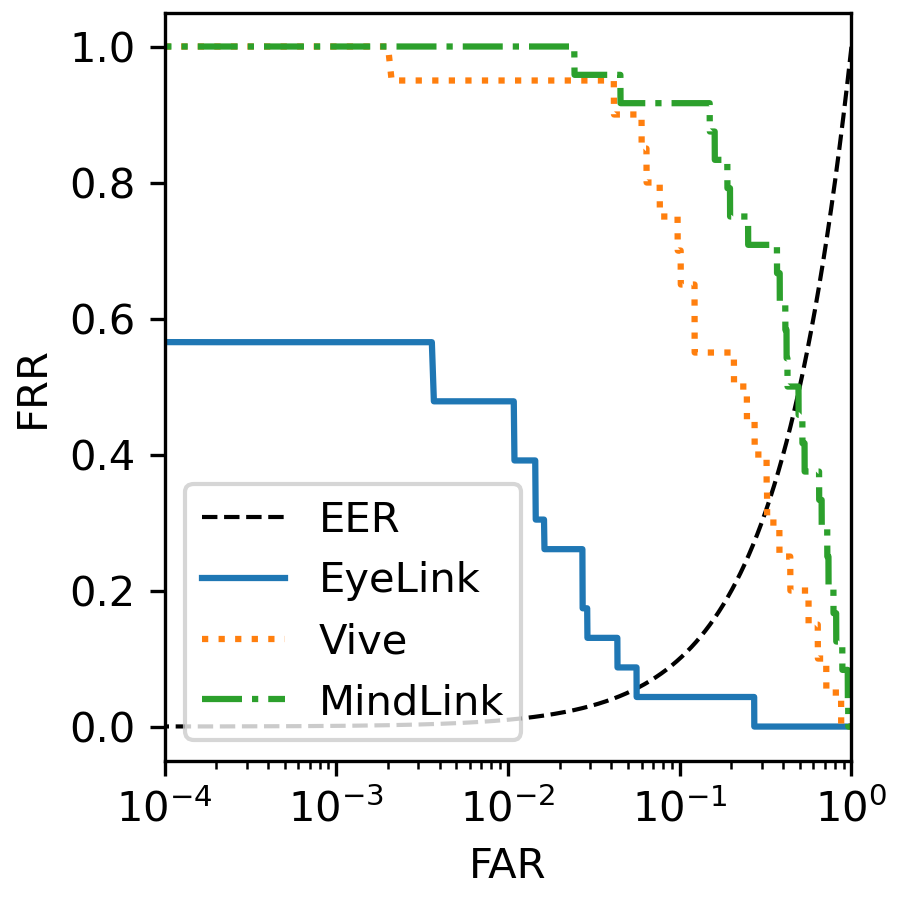}
    \includegraphics[width=0.4\linewidth]{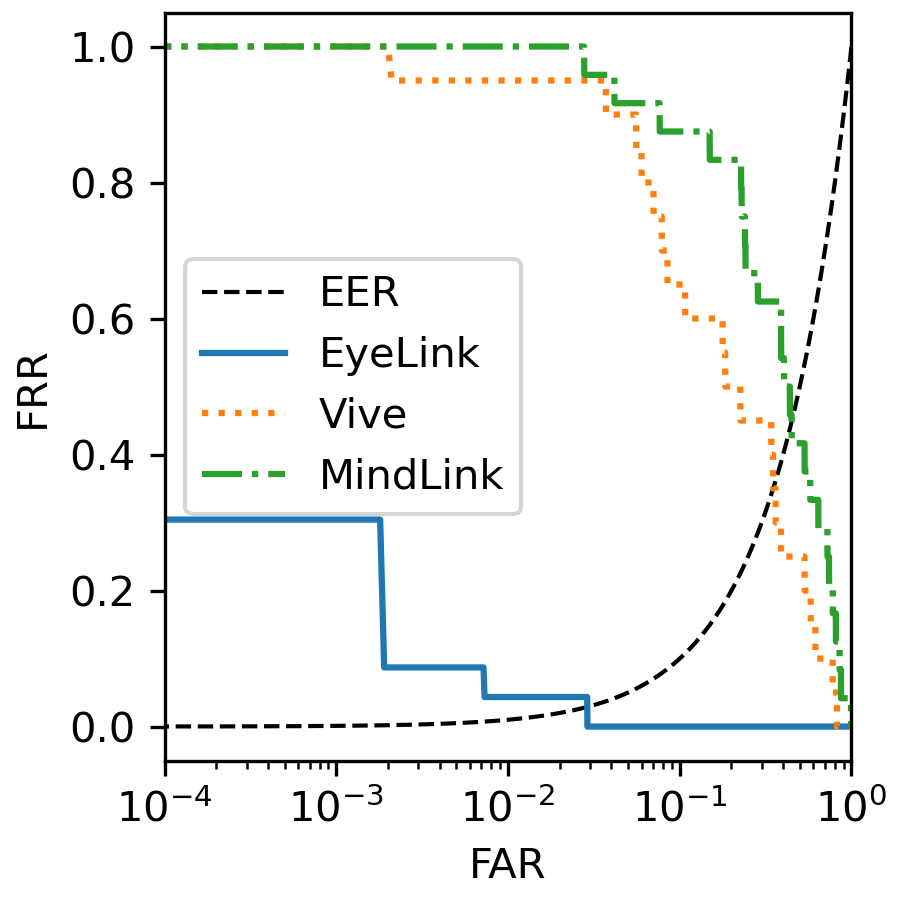}
    \caption{ROC curves for biometric verification on 5-second enrollment (left) and 60-second enrollment (right) across the three devices.}
    \label{fig:roc}
\end{figure}

We observe similar results when evaluating verification performance on 60-second embeddings, as shown in Table~\ref{tab: verification-results-60s}.
Biometric verification performance generally improves when the authentication system is provided with longer sequences of eye movement data---we observe that EER scores for the EyeLink and MindLink improve relative to the 5-second embeddings.
The only exception to this observation is the Vive authentication set, but we note that this degradation in EER likely occurs because many subjects in the Vive data have fewer than 60 seconds' worth of reading data available for analysis, meaning that the resulting embeddings do not represent a full 60 seconds of eye movement.
Overall, it appears that EKYT is most capable of achieving high biometric verification performance on authentication sets that are similar to the data used for enrollment, and is less capable of correctly distinguishing genuine matches from imposters when authenticating with data from different devices.
The ROC curves generated for each verification setting are displayed in Figure~\ref{fig:roc}.

\begin{table}
\begin{center}
\begin{tabular}{ccc}
\hline
Enrollment & Authentication & EER (\%) $\downarrow$\\
\hline
\multirow{3}{*}{EyeLink TEX} & EyeLink RAN & 5.61 \\
 & Vive TEX & 32.23 \\
 & MindLink TEX & 49.13 \\
\hline
\end{tabular}
\end{center}
\caption{Biometric verification results when enrolling and authenticating on 5 seconds of data.}
\label{tab:verification-results-5s}
\end{table}

\begin{table}
\begin{center}
\begin{tabular}{ccc}
\hline
Enrollment & Authentication & EER (\%) $\downarrow$\\
\hline
\multirow{3}{*}{EyeLink TEX} & EyeLink RAN & 2.89 \\
 & Vive TEX & 34.99 \\
 & MindLink TEX & 44.79 \\
\hline
\end{tabular}
\end{center}
\caption{Biometric verification results when enrolling and authenticating on 60 seconds of data.}
\label{tab: verification-results-60s}
\end{table}

\subsection{Cross-Platform Identification}
Tables~\ref{tab:identification-results-5s} and~\ref{tab:identification-results-60s} show identification results across devices when employing 5- and 60-second embeddings, respectively.
Using EyeLink data for authentication produces an IR of 73.91\% for 5-second embeddings and 95.65\% for 60-second embeddings, meaning that 16 and 19 of the 21 subjects were correctly identified.
EKYT exhibits cross-platform identification rates that are slightly above chance levels when authenticating on the Vive data; we observe identification rates of 5\% and 15\% for the 5-second and 60-second embeddings, respectively.
While identification performance using 5-second embeddings produces chance level-performance, we observe that EKYT is still able to produce above-chance identification rates with 60-second embeddings. 
Similar to the verification setting, EKYT demonstrates no ability to re-identify users using the MindLink data as the authentication set when using either 5- or 60-second embeddings, producing an IR of 0.0\% in both cases.

\subsection{Device vs Subject Effects on Embedding Space}
One possible explanation for the observed discrepancy in biometric performance between devices is that the difference in data quality between devices are more prominent than the similarities found in subject-specific characteristics when enrolling and authenticating on different devices.

\begin{figure}
     \centering
     \begin{subfigure}[b]{0.35\linewidth}
         \centering
         \includegraphics[width=\linewidth]{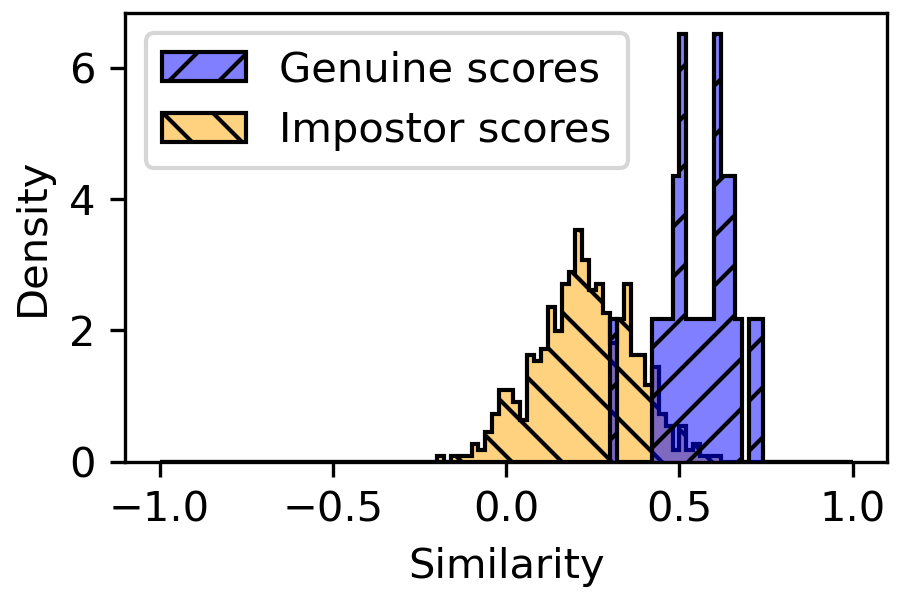}
         \caption{EyeLink, 5-seconds.}
         \label{fig:sim_el_5s}
     \end{subfigure}
      \begin{subfigure}[b]{0.35\linewidth}
         \centering
         \includegraphics[width=\linewidth]{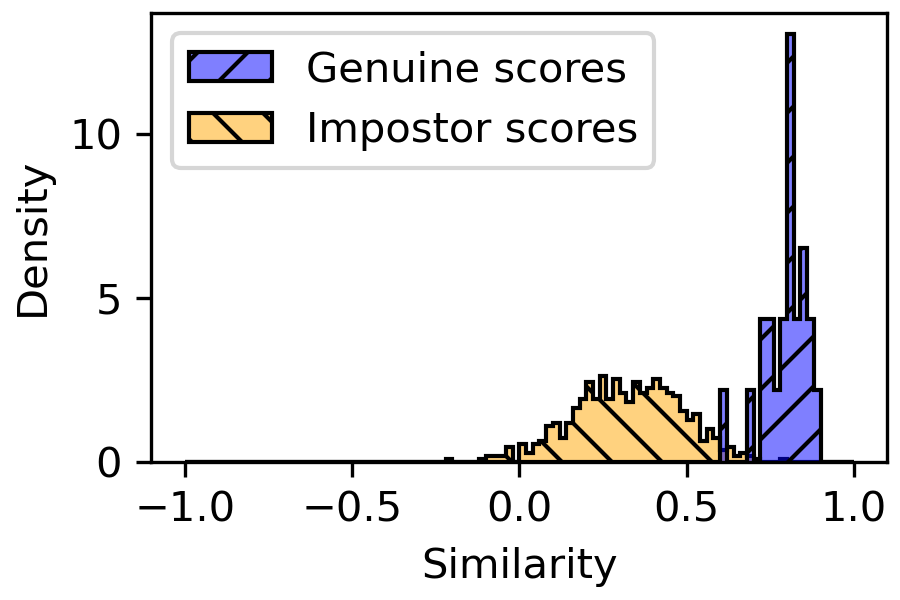}
         \caption{EyeLink, 60-seconds.}
         \label{fig:sim_el_60s}
     \end{subfigure}
     \hfill
     \begin{subfigure}[b]{0.35\linewidth}
         \centering
         \includegraphics[width=\linewidth]{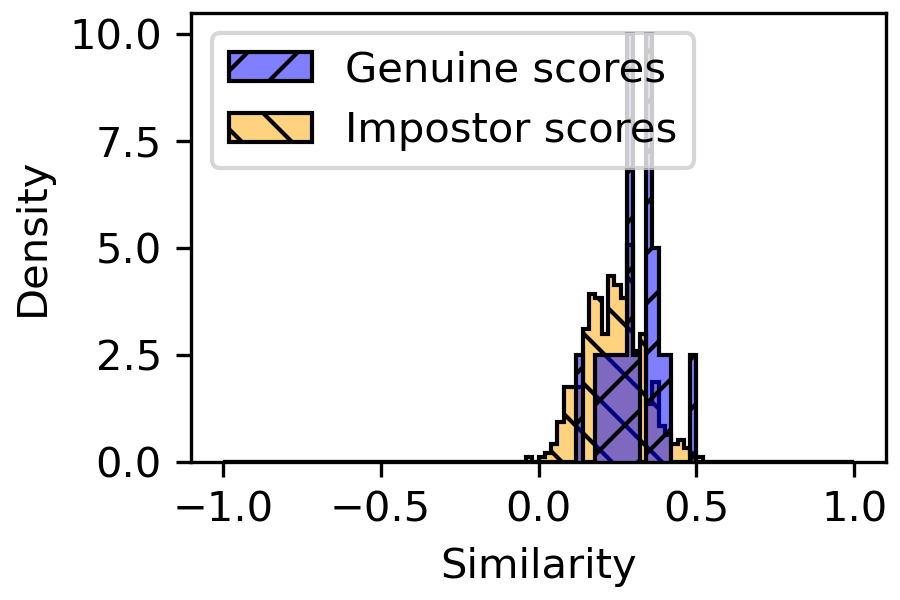}
         \caption{Vive, 5-seconds.}
         \label{fig:sim_vi_5s}
     \end{subfigure}
     \begin{subfigure}[b]{0.35\linewidth}
         \centering
         \includegraphics[width=\linewidth]{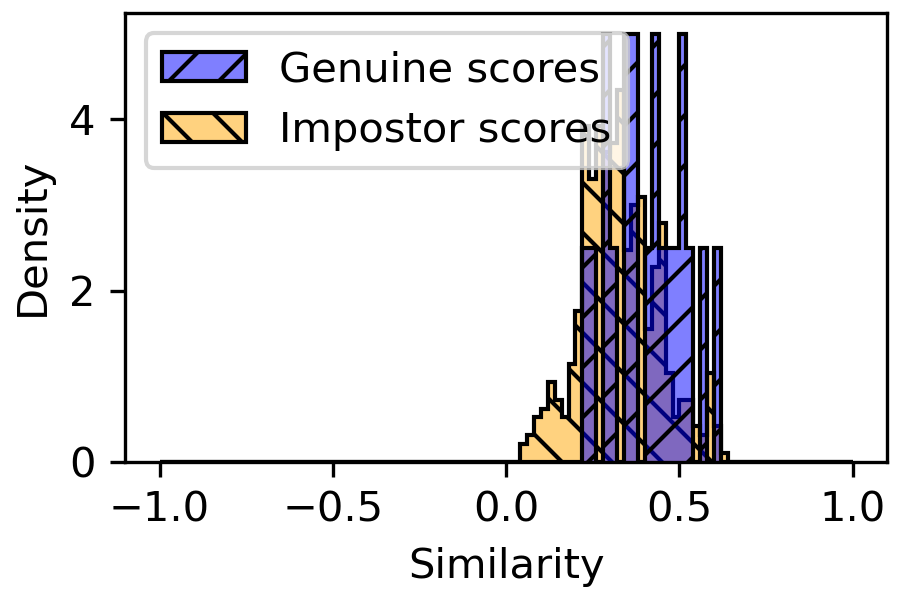}
         \caption{Vive, 60-seconds.}
         \label{fig:sim_vi_60s}
     \end{subfigure}
     \hfill
     \begin{subfigure}[b]{0.35\linewidth}
         \centering
         \includegraphics[width=\linewidth]{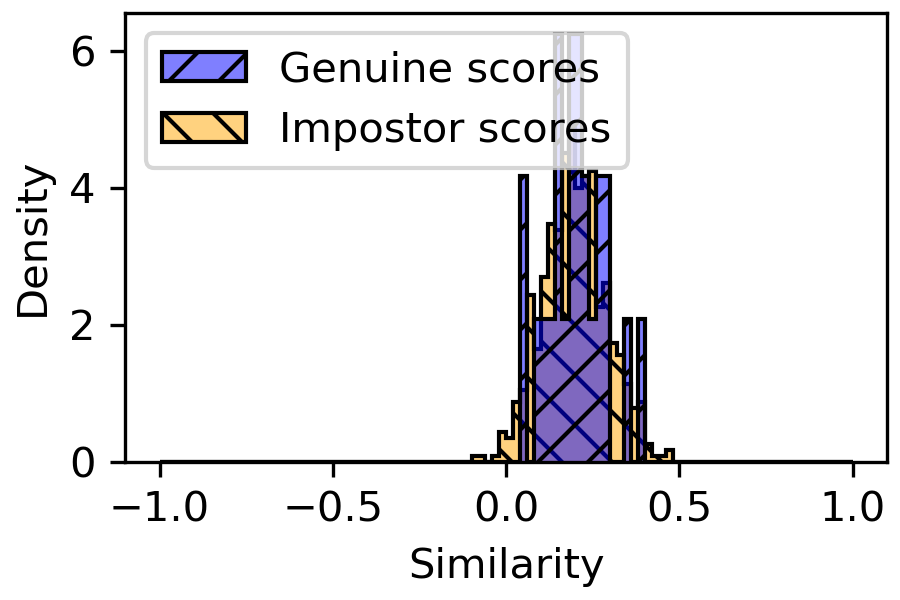}
         \caption{MindLink, 5-seconds.}
         \label{fig:sim_ml_5s}
     \end{subfigure}
     \begin{subfigure}[b]{0.35\linewidth}
         \centering
         \includegraphics[width=\linewidth]{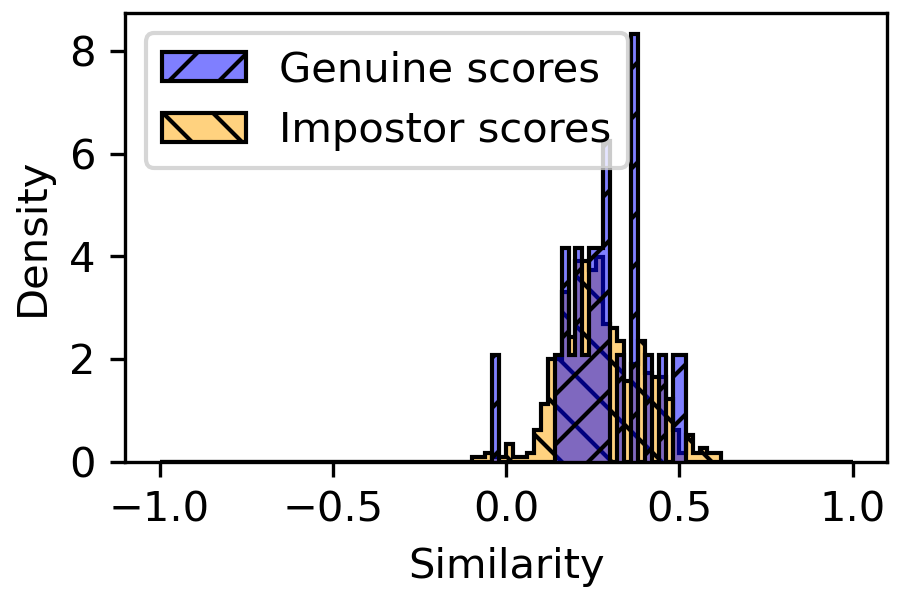}
         \caption{MindLink, 60-seconds.}
         \label{fig:sim_ml_60s}
     \end{subfigure}
        \caption{Similarity distributions for genuine and imposter subjects for embeddings created with 5 seconds and 60 seconds of data.}
        \label{fig:sim-dist}
\end{figure}

Figure~\ref{fig:sim-dist} shows the distribution of similarity scores for genuine and imposter embeddings for each experimental setting. 
The more separated the genuine and imposter distributions are, the better biometric performance the model achieves.
The similarity score distributions when authenticating with EyeLink data (Figures~\ref{fig:sim_el_5s} and~\ref{fig:sim_el_60s}) are very well separated; the overall distribution of all similarity scores is distinctly bimodal, with little overlap between the genuine and imposter distributions.
On the other hand, similarity distributions when authenticating with Vive (Figures~\ref{fig:sim_vi_5s} and~\ref{fig:sim_vi_60s}) and MindLink (Figures~\ref{fig:sim_ml_5s} and~\ref{fig:sim_ml_60s}) data feature notably less separation between genuine and imposter scores.
Because the genuine subject embeddings from these devices produce lower similarity scores, there is significant overlap between the genuine and imposter similarity score distributions.
This makes it harder for the model to distinguish between genuine and imposter subjects, which in turn contributes to the poor biometric performance observed.

\begin{figure*}
    \centering
     \begin{subfigure}[b]{0.4\linewidth}
         \centering
         \includegraphics[]{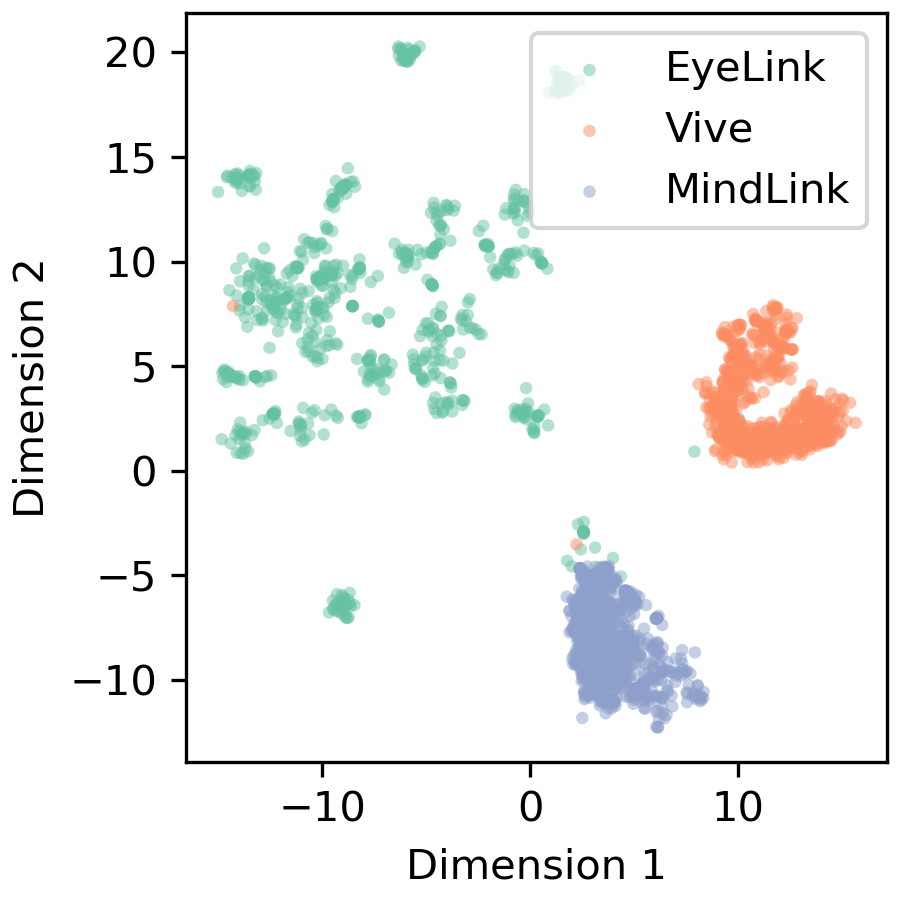}
         \caption{Embeddings labeled by device (N=3).}
         \label{fig:dense-device}
     \end{subfigure}
     \hfill
     \begin{subfigure}[b]{0.4\linewidth}
         \centering      \includegraphics{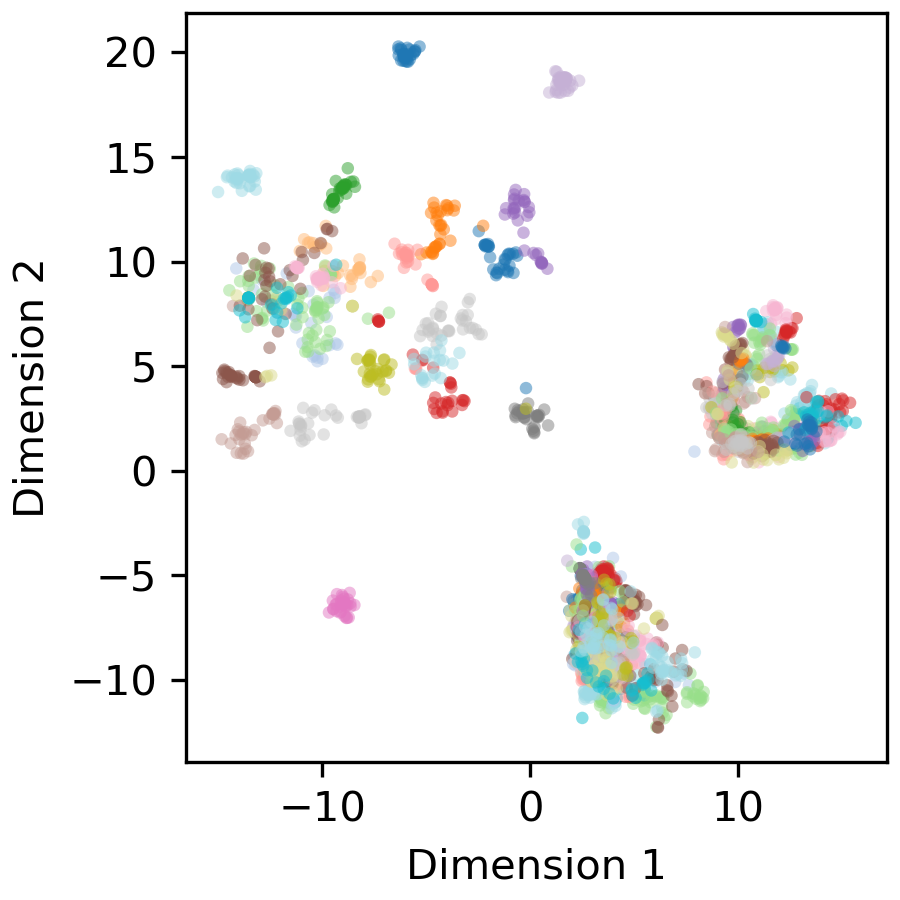}
         \caption{Embeddings labeled by subject (N=21).}
         \label{fig:dense-subject}
     \end{subfigure}
    \caption{A DensMap visualization of the embedding space for all 60-second embeddings generated by EKYT.}
    \label{fig:dense-map}
\end{figure*}

Figure~\ref{fig:dense-map} shows a DensMap~\cite{Narayan2020} visualization of the embedding space for the embeddings produced by EKYT.
This embedding space features three distinct, well-separated clusters corresponding to device (Figure~\ref{fig:dense-device}), and are more loosely concentrated by subject identity within these primary groups (Figure~\ref{fig:dense-subject}).
While this visualization shows that EKYT can meaningfully distinguish between subject embeddings within a device, it also suggests that device type has a much stronger influence over the shape of the embedding space than subject identity.

\begin{table}
\begin{center}
\begin{tabular}{ccc}
\hline
Enrollment & Authentication & IR (\%) $\uparrow$\\
\hline
\multirow{3}{*}{EyeLink TEX} & EyeLink RAN & 73.91 \\
 & Vive TEX & 5.00 \\
 & MindLink TEX & 0.00 \\
\hline
\end{tabular}
\end{center}
\caption{Biometric identification results when enrolling and authenticating on 5 seconds of data.}
\label{tab:identification-results-5s}
\end{table}

\begin{table}
\begin{center}
\begin{tabular}{ccc}
\hline
Enrollment & Authentication & IR (\%) $\uparrow$\\
\hline
\multirow{3}{*}{EyeLink TEX} & EyeLink RAN & 95.65 \\
 & Vive TEX & 15.00 \\
 & MindLink TEX & 0.00 \\
\hline
\end{tabular}
\end{center}
\caption{Biometric identification results when enrolling and authenticating on 60 seconds of data.}
\label{tab:identification-results-60s}
\end{table}

\section{Discussion}
Overall, our results substantiate privacy concerns for cross-platform identity linkage using EMB.
We demonstrate that the state-of-the-art EKYT biometric model achieves above-chance cross-platform verification and identification performance for at least one of the devices examined.
We also observe that eye tracking signal quality may be an obstacle to EKYT's ability to reliably link subject identities between devices, which may be endemic to our choice of biometric model.

EKYT was trained to extract meaningful biometric features from high-quality data captured by a research-grade eye tracker, the EyeLink 1000.
Our results suggest that the features that EKYT learned to extract may not generalize well to a variety of eye tracking signal qualities, as biometric performance degrades significantly when authenticating with data from lower quality eye tracking devices.
However, EKYT meaningfully re-identified subjects using data from the Vive, whose eye tracking signal quality is significantly closer to the EyeLink's than the MindLink's.
While better biometric authentication results could have been achieved by fine-tuning the model with data from all three devices, our threat model assumes that an adversary is not capable of accessing or altering the biometric model to increase their chance of successfully infiltrating the target system.
Because the objective of this study is to establish a baseline of the risk for cross-platform re-identification, altering EKYT to maximize biometric performance across all three devices is outside the scope of our study.

Our investigation first confirms that EKYT achieves similar verification and identification performance on the EyeLink as EKYT's original evaluation.
Lohr and Komogortsev~\cite{Lohr2022_EKYT} originally achieve an EER of 3.5\% and an IR of 91\% when using 5-second embeddings under a similar evaluation paradigm as Tables~\ref{tab:verification-results-5s} and~\ref{tab:identification-results-5s}, meaning that our investigation produces only a nominal degradation in biometric performance relative to the proportion of subjects that are incorrectly verified or identified.

We also observe above-chance verification rates when using the Vive data as the authentication set, indicating that subject-specific characteristics that are shared in both subsets of eye tracking data can be meaningfully extracted and linked.
Identification rates when using the Vive data do not meaningfully exceed chance-level performance when using 5-second embeddings.
However, we observe above-chance identification rates when using 60-second embeddings from the Vive data, which indicates that meaningful biometric identification across devices is possible given a sufficient volume of data.

When using the MindLink data for authentication, we observe biometric  verification performance approaching chance levels, meaning that data from the MindLink did not enable cross-platform identity linkage.
Using the MindLink for biometric identification produces similar results---when using both 5-second embeddings and 60-second embeddings, the MindLink data produced a 0\% identification rate.
This finding is particularly interesting, as the EyeLink and MindLink data used for enrollment and authentication were collected simultaneously from the same subjects at the same time.
Theoretically, the only differences in the data from these two devices are the eye that data was collected from and differences in eye tracking signal quality produced by each device's gaze estimation pipeline.
It is possible that, although the pre-trained EKYT model can identify unique subject characteristics within unseen devices, the differences in eye tracking signal quality between devices hinders its ability to successfully associate embeddings belonging to the same identity across two devices.

The distribution of similarity scores shown in Figure~\ref{fig:sim-dist} further illustrates the discrepancy in performance between devices.
The low degree of separation between genuine and imposter authentication attempts contributes to the relatively poor biometric performance for the Vive and the MindLink.

We further illustrate the behavior of the model by visualizing the embedding space for all embeddings in Figure~\ref{fig:dense-map}.
While we would expect EKYT to effectively distinguish EyeLink data from non-EyeLink data, we did not expect it to also clearly distinguish between data collected from the Vive and the MindLink, despite never being exposed to data from either device during training.

These results suggest that the relative differences in eye movement signal quality between devices affects the model's ability to meaningfully link identities across devices. 
The Vive data and the EyeLink data are relatively more similar in terms of spatial accuracy and precision than the MindLink (Table~\ref{tab:data}), which may contribute to the better biometric performance observed when using Vive data for authentication.
It is an open research question whether the authentication model additionally identifies artifacts in the data specific to a device's gaze estimation pipeline, rather than the signal quality itself.
This raises the question of whether an EMB authentication system would be susceptible to cross-platform identity linkage if the two different eye tracking devices featured different levels of signal quality, but were created by the same manufacturer.

\section{Limitations}
We acknowledge that this study has some limitations that may have affected overall biometric performance, largely stemming from the availability of paired eye movement data.

Firstly, the data set employed for this investigation contains a very small number of subjects.
Paired eye movement data is limited in availability and can be costly to produce; if a larger data set of paired eye movement emerges, future work should re-evaluate the results obtained in this investigation with a larger cohort of subjects.
Secondly, the data from each device used in this investigation were collected from different eyes; the EyeLink portion was captured from the left eye, while the MindLink and Vive portions were captured from the right.
While gaze positions between the two eyes are typically highly correlated~\cite{Aziz2022}, using data streams from different eyes may affect overall biometric performance when using eye movement tasks like reading, which involve left-to-right eye movement patterns.
Additionally, downsampling the EyeLink and MindLink data to 250 Hz could have also affected biometric performance.
However, we suggest that the potential impact of downsampling data on our results is minimal, as we successfully replicate the biometric performance observed in the EKYT's original investigation using the EyeLink portion of the data.
A device's native eye tracking signal quality likely has a larger effect on biometric performance than downsampling does.

\section{Conclusions and Future Work}
This work explores the privacy implications of ubiquitous eye tracking, a topic that is often discussed but seldom investigated in eye tracking literature.
Specifically, we quantify a modern EMB-based authentication model's ability to perform cross-platform identity linkage, where a user's identity is connected across different devices that use eye tracking technology.
We show that a modern EMB system can achieve above-chance biometric authentication performance when using data from two different eye tracking devices, thereby substantiating concerns for user privacy in environments with ubiquitous eye tracking. 

This investigation advances a fundamental understanding of the privacy risk of cross-platform identity linkage for eye tracking.
By demonstrating an appreciable risk for cross-platform identity linkage via eye tracking, our study can guide the development of practical defenses and privacy preserving techniques that prevent both EMB systems and eye tracking-enabled platforms from being exploited to violate user privacy.
Future work in eye tracking privacy can shed further light on this privacy risk by involving a larger population of users across a variety of eye tracking devices of varying levels of similarity.

\section{Acknowledgements}
This material is based upon work supported by the National Science Foundation Graduate Research Fellowship under Grant No. DGE-1840989. Any opinion, findings, and conclusions or recommendations expressed in this material are those of the authors(s) and do not necessarily reflect the views of the National Science Foundation.

\bibliographystyle{plain}
\bibliography{references}

\end{document}